\documentclass[aps,preprint]{revtex4}%
\usepackage{amssymb}
\usepackage{amsfonts}
\usepackage{amsmath}
\usepackage{graphicx}%
\providecommand{\U}[1]{\protect\rule{.1in}{.1in}}
\begin{document}
\title{Critical behavior of Ising spins in a tridimensional percolating nano system
with noninteger fractal dimension.}
\author{Gennadiy Burlak}
\affiliation{Centro de Investigaci\'{o}n en Ingenier\'{\i}a y Ciencias Aplicadas,
Universidad Aut\'{o}noma del Estado de Morelos, Cuernavaca, Mor. Mexico.}
\keywords{percolation, Ising model, critical phenomenon}
\pacs{05.50.+q, 75.10.Hk}

\begin{abstract}
In an artificial 3D percolation nano medium, the clusters filled by the Ising
magnets give rise to a topologically nontrivial magnetic structure, leading to
new features of the ferromagnetic phase transition without an external
magnetic field. In such an inhomogeneous system, the standard Ising model is
strongly modified by the spatial percolation cluster distribution. We found
numerically that at percolation occupation probability $p<1$ far from the
percolation threshold $p_{c3D}$, the magnetization shows
ferromagnetic-paramagnetic phase transition with the transition temperature
$T_{c}$ depending considerably on the probability $p$. We provide numerical
evidence that in vicinity $p_{c3D}$\ the dependence $T_{c}(p)$ is affected by
the noninteger fractal dimension $D_{H}(p)$ of the incipient percolation
spanning cluster.

\end{abstract}
\maketitle

\section{Introduction}

The Ising model with no external magnetic field possesses by the critical
behavior, when below the critical temperature the phase transition from
paramagnetic state to ferromagnetic state occurs. However, do the critical
properties of the three dimensional (3D) Ising model still exist for a
noninteger dimension case with a fractal (Hausdorff) dimension of $D_{H}<3$?
(Let us remind that the fractal dimension, $D_{H}$, is a statistical quantity
that gives an indication of how completely a fractal appears to fill space, as
one zooms down to finer and finer spatial scales\cite{Wiki}) To answer this
question, we study the nano structure of a spanning cluster in 3D embedded
percolation environment for the creation an object with a fractional
dimension. The spanning cluster in such a situation provides the connection
between the input and output (opposite sides) of the sample. As a result, the
opportunity to incorporate the spins into such an opened nano structure
becomes possible. Forcing the spin ensemble through such a percolation medium
allows creating a network of spins, leaving other parts of the system
unchanged. Since for 3D geometry the fractal dimension of the incipient
spanning cluster is $D_{H}\approx2.54$ (see e.g. \cite{Isichenko:1992a},
\cite{Kroger:2000a}), in principle, this allows studying experimentally the
properties of phase transition in the Ising model in geometry with noninteger dimension.

The original idea to study critical behavior for non-integer space by use of a
fractal lattice is due to Gefen, Mandelbrot and Aharony\cite{YuvalGefen:1980a}%
. The spin model chosen is the Ising model, where spins are classical
variables taking the values $\pm1$. The idea suggested by Gefen et. al. was to
construct a space domain with noninteger dimension via a fractal lattice (the
Sierpinski carpets), and then study its critical behavior as a function of the
noninteger dimension. However, in which way could such a critical behavior be
studied experimentally in spin- and optoelectronics?

The Ising model near criticality in 2D and 3D dimensions has been studied
numerically via Monte Carlo methods by Cambier and
Nauenberg\cite{Cambier:1986a}. They  considered the system closely below the
critical temperature $T_{c}$ and studied the cluster size distribution. Using
such definition of critical clusters, Coniglio\cite{AntonioConiglio:1989a} was
able to compute the fractal dimension of these clusters, by mapping the Potts
model to the Coulomb gas. The Coniglio-Klein clusters of the Ising model in 3
dimensions have been investigated by Wang and Stauffer\cite{Wang:1990a} using
Monte Carlo simulations.

Let us recall that for fractals involving a random element, one speaks about a
statistical self-similarity, meaning equivalent, after the proper rescaling,
statistical distributions characterizing the geometry of a part and of the
whole fractal. It's worth to note that such a compound system (clusters with
spins) can be created in the experiment if to fill the voids of percolating
clusters by nano-particles of a ferromagnetic (magnetorheological) fluids that
recently were synthesized\cite{Kose:2009a}-\cite{Kikuraa:2007a}. The
registration of incipient percolation cluster by optical methods was recently
considered in Ref.\cite{Burlak:2009a}, while the specific dynamical field
properties of radiated nanoemitteres incorporated in a percolating cluster are
studied in Ref.\cite{Burlak:2009b}.

In a percolation medium filled by the spins, there are two distinct physically
important order parameters that are defined by a geometrical configuration of
the nearest neighbors in such a compound system. The first (percolating
$P_{3D}$) order parameter is defined by the occupation probability of the
lattice cells $p$. The percolation structure normally creates a hollow fractal
network of channels that can be filled by a liquid, gas, conductivity
electrons, etc.

In this paper we consider the properties of a 3D percolation medium where the
percolation channels are filled by the spins. The second order parameter
(average magnetization $M$) depends on the temperature $T$ . We numerically
show that the critical properties of such an extended Ising model (EIM) depend
on the percolation occupation probability $p$ and the temperatures $T$ as
well. If the probability $p$ is close to $1$ (far from the percolation
criticality), then the percolation clusters expand throughout the system. In
such a situation, the critical properties of the EIM mainly coincide with
well-known unbounded case\cite{Isichenko:1992a}, \cite{Kroger:2000a}. However,
for a smaller $p$ (closer to the percolation phase transition), the system
becomes quite inhomogeneous, and does not yet possess a translation symmetry.
As our simulations have shown, even in such a situation, the ferromagnetic
phase transition in EIM still exists; however, the critical temperature
$T_{c}$ acquires considerable dependence on the occupation probability $p$. In
this case, the dependence $T_{c}(p)$ is affected by the noninteger fractal
dimension of the incipient percolation spanning cluster.

The properties of spins system in $3D$ disordered mediums are an area of
active research\cite{Kormann:2010a}-\cite{Middleton:2002}. Nevertheless, the
critical properties of the Ising system in a network of percolation $3D$
clusters are still studied insufficiently. However, the phase transition of
spontaneous magnetization driven by concentration $p$ of the pores and spins
(incorporated into such voids) is very attractive for various applications of
quantum spin- and optoelectronics. As far as the author is aware, the critical
properties of Ising spins incorporated into a 3D percolation medium has poorly
been considered thus far, though it is a logical extension of previous work in
this area.

This paper is organized as follows. In Sec. II, the basic equations and our
numerical scheme for studying the structure of $3D$ percolation clusters is
discussed. In Sec. III, we study the ferromagnetic phase transition of Ising
magnets incorporated into the spanning cluster, which is considered in Sec.
II. Discussion and conclusions from our results are found in the last Section.

\section{Basic equations and numerical algorithm.}

First, we study the geometrical properties of percolation clusters in a cubic
lattice embedded in a 3-dimensional (3D) medium. It is assumed that the
probability $p$ for each site of the lattice to be "occupied" is given, and we
investigate the spatial distribution of the resulting clusters over sizes and
other geometrical parameters. \ We recall that a cluster here means a
conglomerate of occupied sites, which communicate via the nearest-neighbor rule.%

%TCIMACRO{\FRAME{ftbpFU}{5.8937in}{4.4373in}{0pt}{\Qcb{(Color online). Spatial
%distribution of the spanning cluster. Supercritical concentration of defects
%($p=0.317$) is taken slightly above threshold $p_{3D}$. Only clusters are
%shown, contacting with the input side of a material opened for the spin flow.
%We observe that a spanning cluster which spans the sample completely from
%input up to output sides has arisen. See details in the text.}}%
%{\Qlb{Pic_thrperc}}{fig1.eps}{\special{ language "Scientific Word";
%type "GRAPHIC";  maintain-aspect-ratio TRUE;  display "USEDEF";
%valid_file "F";  width 5.8937in;  height 4.4373in;  depth 0pt;
%original-width 5.834in;  original-height 4.3863in;  cropleft "0";
%filename 'Fig1.eps';file-properties "XNPEU";}}}%
%BeginExpansion
\begin{figure}
[ptb]
\begin{center}
\includegraphics[
height=4.4373in,
width=5.8937in
]%
{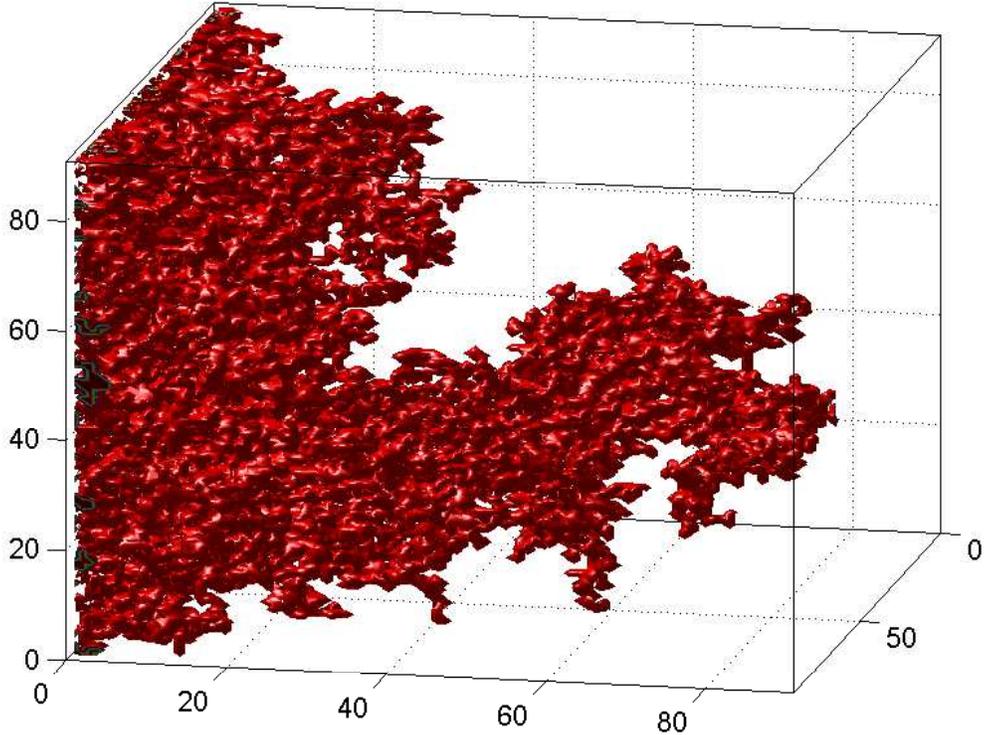}%
\caption{(Color online). Spatial distribution of the spanning cluster.
Supercritical concentration of defects ($p=0.317$) is taken slightly above
threshold $p_{3D}$. Only clusters are shown, contacting with the input side of
a material opened for the spin flow. We observe that a spanning cluster which
spans the sample completely from input up to output sides has arisen. See
details in the text.}%
\label{Pic_thrperc}%
\end{center}
\end{figure}
%EndExpansion

In our algorithm, the connectivity of cluster voids (pores) is initially
examined locally, so that two pores produce a cluster if they have at least
one edge in common. In such an approach, the spanning occurs when the size of
the largest (spanning) cluster reaches the size of the system. The rest of the
cluster is referred to as a collection of "dead" or "dangling" ends. In a
large enough sample, the internal topology is changed substantially at the
critical concentration of the percolation voids $p_{c3D}$. We have found that
such an intrinsic topological structure affects the dimensionality of the
percolation transition; as a result, the connectivity of a percolating network
acquires the form of a fractal, see Fig.\ref{Pic_thrperc}. It is worth noting
that in the three-dimensional case, no percolation threshold is known exactly;
only numerical data are available\cite{Isichenko:1992a}. Fig.\ref{Pic_thrperc}
shows the incipient spanning cluster for the supercritical probability
$p\gtrsim p_{3D}=0.311$ when the spanning $3D$ cluster emerges for the first
time forming a fractal network of voids.

In general\textbf{,} the analysis of such a compound system requires quite
long computations. The first step deals with the identification of the
spanning cluster $P_{3D}$ as a function of ocupations probability $p$. In this
step, the percolation system on a lattice is simulated by a random variable
$s$ that accepts only values $1$ or $0$ (with probability $p$) that defines
whether or not a cell is a percolating one. The percolation order parameter
$P_{3D}$ is defined as the ratio of the number of cells belonging to the
spanning cluster to the general number of cells. Obviously that $P_{3D}$ is
distinct from zero only when exceeding the threshold concentration $p\gtrsim
p_{3D}$. Corresponding dependencies are shown in Fig.\ref{Pic_Fig1}. In the
second step, the properties of spins incorporated into the percolation
structure (known from the first step) are evaluated with the use of the Monte
Carlo technique (see \cite{Stauffer:1992}, \cite{Robert:2004a} and references
therein). In this step, all values $1$ in conducting cells are replaced (with
probability $0.5$) by $+1$ or $-1$ instead; correspondingly parallel or
antiparallel spin state in such a cell is generated.

The resulting average magnetization $M$ will be defined not only by the
temperature $T$ (more details are given below, see Eq.(\ref{metrop})), but
also by the occupation probability $p$. We recall that the nature of a
percolation problem as a critical phenomenon is characterized by the critical
exponents $\beta$ and $\nu$. In theory, the exponent $\beta$ is related to the
intensity of the order parameter $P_{t}$ (we supply $P_{t}$ by index $t$ to
separate this one from the numerically evaluated quantity $P_{3D}$) as%
\begin{equation}
P_{t}=(p-p_{3D})^{\beta}\text{, }p>p_{3D}\text{, \ and }P_{t}=0\text{ for
}p\leq p_{3D\text{,}}\label{P_t}%
\end{equation}
and the exponent $\nu$ is related to the correlation length $\xi$ as%
\begin{equation}
\xi\approx\left\vert p-p_{3D}\right\vert ^{-\nu}\text{, where }p\simeq
p_{3D}.\label{ksi}%
\end{equation}
The latter is a measure of the range over which fluctuations in one region of
space are correlated with those in another region. Two points which are
separated by a distance larger than the correlation length $\xi$ will each
have fluctuations which are independent, that is, uncorrelated. Thus the
divergence of the correlation length $\xi$ at the critical point means that
points being very far from each other become correlated. The exponents $\beta
$\ and $\nu$ are ones of the standard set of critical
exponents\cite{Domb:2000a} that govern the singular behavior of different
quantities near the critical point. These exponents depend only on the
dimension of the space and not on the type of lattice or the kind of
percolation problem.

In two dimensions ($d=2$), these $\beta$\ and $\nu$ are known
analytically\cite{denNijs:1979a}, \cite{Pearson:1980a}, namely, $\nu=4/3$ and
$\beta=5/36$. For $d=3$, only numerical estimates are available: $\nu
\simeq0.90$ and $\beta\simeq0.40$ (see Table III in Ref.\cite{Isichenko:1992a}%
). It has been established that critical exponents $\nu$ and $\beta$, as well
as all others, do not depend on the kind of lattice (e.g., square, triangular,
etc.) or the kind of percolation problem (site or bond). The corresponding
power exponents are structurally stable; that is, they are unchanged by a
small perturbation of the lattice model or the mapping, respectively, see
\cite{Isichenko:1992a} and references therein. From the finite size scaling
theory\cite{Isichenko:1992a}, \cite{Stauffer:1992} various values of $P_{t}$
at given values of $p$ and $L$ can be fitted by a single scaling function
$F(x)$ (that depends only on a single variable, but is not otherwise given
explicitly by the theory\cite{Fisher:1974}, \cite{Grimmett:1999a}) that leads
to strong dependence of order parameter on the values of $\beta$ and $\nu$.
The dependence $p_{3D}$ on the value $L$ is rather weak.

\section{Numerical simulations of the 3D cluster structure}%

%TCIMACRO{\FRAME{ftbpFU}{5.8807in}{4.4227in}{0pt}{\Qcb{\QTR{group}{(Color
%online). The order parameter $P_{3D}(p)$, fractal dimension $D_{H3D}/d3$ of
%the spanning percolation cluster, and critical temperature $T/T_{c3D}$ for
%Ising spins (placed in such a cluster), as function of the occupation
%probability $p$. To compare different dependencies, the following
%normalizations are used: $d_{3}=3$, $T_{c0}=4.54$. The size of the system is
%$L^{3}=140^{3}$; $dt(p)$ is the time interval required to calculate
%corresponding value $P_{3D}(p)$. The percolation arises close to the critical
%value $p_{3D}\thickapprox0.311$, where the fractal dimension becomes
%noninteger $D_{H3D}=2.54$, and the value $dt(p)$ has a high peak. See details
%in text.}}}{\Qlb{Pic_Fig1}}{fig2.eps}{\special{ language "Scientific Word";
%type "GRAPHIC";  maintain-aspect-ratio TRUE;  display "USEDEF";
%valid_file "F";  width 5.8807in;  height 4.4227in;  depth 0pt;
%original-width 5.8219in;  original-height 4.3708in;  cropleft "0";
%filename 'Fig2.eps';file-properties "XNPEU";}}}%
%BeginExpansion
\begin{figure}
[ptb]
\begin{center}
\includegraphics[
height=4.4227in,
width=5.8807in
]%
{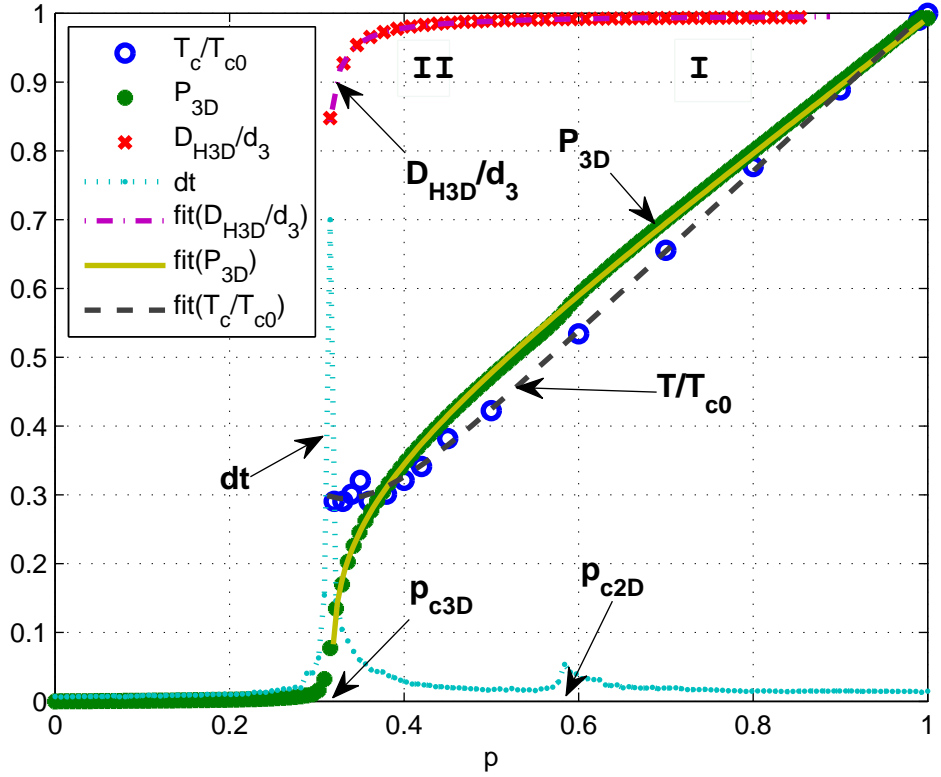}%
\caption{{(Color online). The order parameter $P_{3D}(p)$, fractal dimension
$D_{H3D}/d3$ of the spanning percolation cluster, and critical temperature
$T/T_{c3D}$ for Ising spins (placed in such a cluster), as function of the
occupation probability $p$. To compare different dependencies, the following
normalizations are used: $d_{3}=3$, $T_{c0}=4.54$. The size of the system is
$L^{3}=140^{3}$; $dt(p)$ is the time interval required to calculate
corresponding value $P_{3D}(p)$. The percolation arises close to the critical
value $p_{3D}\thickapprox0.311$, where the fractal dimension becomes
noninteger $D_{H3D}=2.54$, and the value $dt(p)$ has a high peak. See details
in text.}}%
\label{Pic_Fig1}%
\end{center}
\end{figure}
%EndExpansion

Motivated by the essential fractal form of the incipient spanning cluster
shown in Fig.\ref{Pic_thrperc}, we study the dependence of the fractal
dimension of a percolation cluster $D_{H}(p)$ on probability $p$, see
Fig.\ref{Pic_Fig1}. (At calculation $D_{H}$ it was used the approach
\cite{method} that we have adapted for evaluation of the fractal dimension for
3D spanning cluster.) From Fig.\ref{Pic_Fig1} we observe that the behavior of
$D_{H}(p)$ is separated into two well defined parts: $I$ and $II$. In part $I$
for $p>0.5$ (when the spanning clusters expand throughout the system), the
fractal dimension $D_{H}$ increases very slowly from $2.96$ to $3$ (the
dimension of the embedded $3D$ space), see Fig.\ref{Pic_Fig1}. However, in the
part $II$ (critical zone) the value $D_{H}$ increases rapidly from $2.54$ to
$2.96$ when $p$ increases from the critical probability $p_{c3D}=0.311$ to
$0.5$. For our simulation we have used a $3D$ lattice with $L^{3}=N^{3}$ where
$N=80,90,120,140$, but since corresponding curves were very closely placed
only case $N=140$ in Fig.\ref{Pic_Fig1} is shown.

From our numerical data in Fig.\ref{Pic_Fig1} we found quite simple analytical
formula (similar the shifted Boltzmann distribution (with respect to $p$))
that fits the fractal dimension of the spanning cluster:%

\begin{equation}
D_{H}(p)=a\exp(-b/(p-c)),p>c, \label{DH}%
\end{equation}
where $a=2.99$, $b=0.00194$, and $c=0.303$. The behavior of the percolation
order parameter $P_{3D}(p)$\ is fitted with great accuracy (with $\chi
^{2}=2\cdot10^{-5}$) by the expression : $P_{3D}(p)\thickapprox\Psi
(p)=[(p-p_{c3D})^{C}+A]^{B}$, with $0.315<p<1$, and $A=0.024$, $B=0.294$,
$C=3.21$. This differs from the well-known percolation critical dependence
$\left\vert p-p_{c3D}\right\vert ^{-0.15}$\ that is valid only in vicinity
$\left\vert p-p_{c3D}\right\vert \rightarrow0$.

Furthermore, we suppose that magnetic "liquid" (spins) fills up the
percolation structure (from input to output), and as a result the network of
3D Ising magnets is generated. Since the fractal dimension $D_{H}$ of spanning
cluster is noninteger, it is interesting to study the
ferromagnetic-paramagnetic phase transition in such a compound system (spins
in the percolation cluster) at the change of temperature $T$ (see
Eq.(\ref{metrop})) where the energy Hamiltonian has well-known
form\cite{Wang:1990a}, \cite{Kroger:2000a}:%
\begin{equation}
{\mathcal{H}}=-J\sum_{\langle ij\rangle}S_{i}S_{j},\label{eq:H}%
\end{equation}
where $J>0$ (throughout this paper we set $J=1$) is the coupling constant
between nearest-neighbor spins $S_{i}$ and $S_{j}$, where $\langle ij\rangle$
denotes that the sum runs over next-neighbor sites. We are interested, if the
magnetization $M$ in such a percolation fractal system still has critical
property of separation in the ordered and disordered magnetic phases. In
general, the spatial non-uniform percolation effects lead to a complicated
magnetic energy landscape, so the finding of an optimal energy state becomes a
global optimization problem. \ 

The simulations are performed with single spin-flip rates: local changes in
the system are accepted with the probability $\pi$ given by the Metropolis rate,%

\begin{equation}
\pi=\min(1,\exp(-\Delta{\mathcal{H}}/T)),\label{metrop}%
\end{equation}
where $T$ is the temperature and $\Delta{\mathcal{H}}$ is the energy
difference corresponding to the proposed single spin flip\cite{Kroger:2000a}%
,\cite{Janke:2002a}.

\begin{figure}
[ptb]
\begin{center}
\includegraphics[
height=4.4227in,
width=5.8807in
]%
{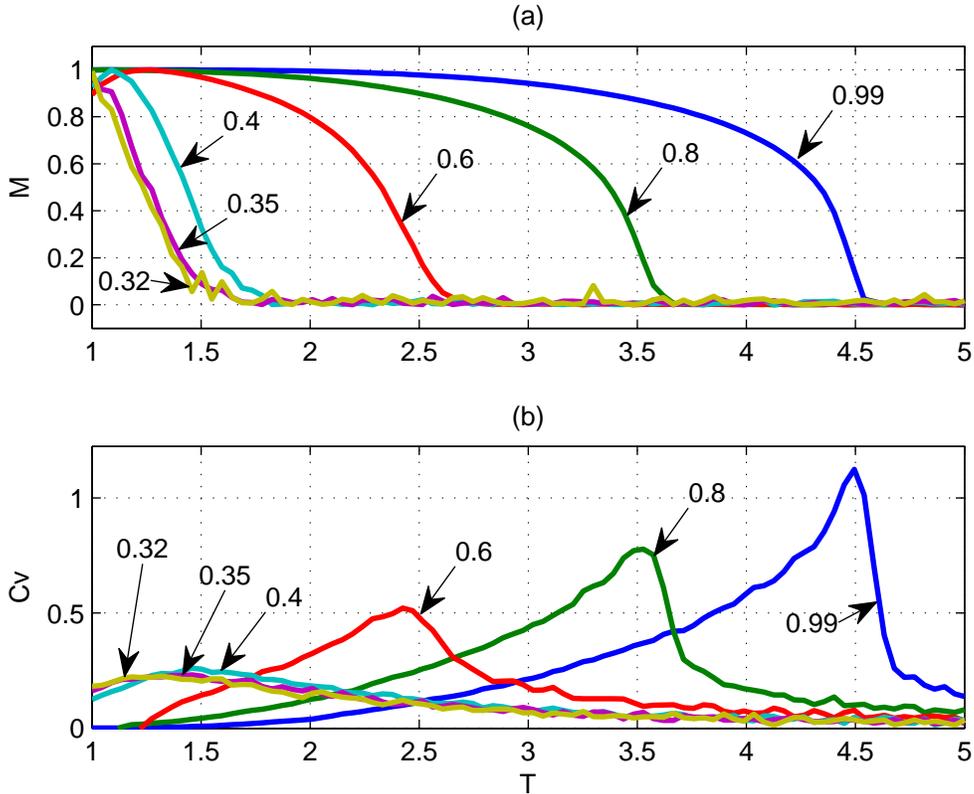}%
\caption{{(Color online) Plot of spontaneous average magnetization $M$ and
specific heat $C_{V}$ corresponding to disordered phase ($T>Tc$) and broken
phase ($T<Tc$) at various probabilities $p$. The continuous curves are a fit
to the data points. We observe that for large $p\leq1$ values, the behavior
$M$ and $C_{v}$ is similar to the unbounded case, but with the shift of the
critical temperature $T_{c}(p)$. }}%
\label{Pic_MCv}%
\end{center}
\end{figure}

Fig.\ref{Pic_Fig1} shows the dependence of the normalized critical temperature
$T_{c}(p)/T_{c0}$ of the phase transition $T_{c}(p)$ on the occupation
probability $p$, where $T_{c0}=4.54$ is the critical Ising temperature for
\ the unbounded $3D$ geometry. (More details are depicted in Fig.\ref{Pic_MCv}%
.) It is worth to note that the Metropolis algorithms shows the critical
slowing down that leads to a dramatic increase of temporal scales in vicinity
of ferromagnetic phase transition. This is due to a high energy penalty
required to flip a single spin (or pair of spins) in a cluster of uniformly
oriented neighbors\cite{Janke:2002a}. To overcome this problem, the cluster
algorithms with non-local dynamics, the Swendsen-Wang\cite{Swendsen:1987a} and
Wolff \cite{Wolff:1989a} approaches commonly used for simulation in the
vicinity of the criticality. As we observe from Fig.\ref{Pic_Fig1} beyond of
the area of phase transition the Metropolis algorithm can be used again for
calculation of that average magnetization.

From numerical data shown in Fig.\ref{Pic_Fig1} we have found the expression
that fits the dependence of the critical temperature as%

\begin{equation}
T_{c}(p)/T_{c0}=p_{1}\exp(-p/p_{2})+p_{3}+p_{4}p, \label{Tc}%
\end{equation}
with $p_{1}=18.44$, $p_{2}=0.059$, $p_{3}=-0.162$, $p_{4}=1.16$. It is
noticeable that in the latter all numerical coefficients $p_{i}$ have a rather
slow dependence on the system size $L$.

As already it was mentioned in the above such an extended Ising system has two
order parameters: the percolation order parameters $P_{3D}$ and the average
magnetization $M$. It is instructively to compare the behavior with respect of
$p$ other quantities that are closely connected with both $P$ and $M$, namely
the fractal dimension of percolation clusters $D_{H}(p)$ (\ref{DH}) and
critical temperature of the ferromagnetic phase transition $T_{c}(p)$
(\ref{Tc}) correspondingly. From Fig.\ref{Pic_Fig1} we observe that in area of
part $I$ (where $D_{H}(p)\simeq3$) the dependence $T_{c}(p)$ is practically
lineal; for such an area ($p\leq p_{c3D}$) in $T_{c}(p)$, the exponential term
$\sim\exp(-p/p_{2})$ is negligibly small. However, in zone $II$, where $D_{H}$
is noninteger, the term $\exp(-p/p_{2})$ sufficiently affects the linear
dependence $T_{c}(p)$: in such a zone $T_{c}$ slowly depends on $p$, and
$T_{c}(p)\sim0.3\cdot T_{c0}$. \ More details of critical behavior $M(T)$ and
specific heat $C_{v}(T)$ are shown in Fig.\ref{Pic_MCv}. From
Fig.\ref{Pic_MCv} (a) we observe that the temperature dependencies $M(T)$ have
well separated disordered phase (at $T>T_{c}$) and ordered (ferromagnetic)
phase (at $T<T_{c}$); the magnetization drops off sharply near the critical
temperature $T_{c}$. Fig.\ref{Pic_MCv} (b) shows that the specific heat
$C_{v}(T)$ has well defined peak at the critical temperature $T_{c}$.
Furthermore, the value of the critical temperature $T_{c}$ strongly depends on
the probability $p$: $T_{c}(p)$ decreases with decreased $p$. We found that
the critical behavior $M(T)$ and $C_{v}(T)$ is similar to the unbounded case,
but with the shift of the critical temperature dependent on probability $p$.
Fig.\ref{Pic_Fig1} shows that for area $I$ fractal dimension $D_{F}$ the of
percolation cluster is nearly equal to the dimension of embedded space
$d_{3}=3$, and $T_{c}(p)/T_{c3D}$ is close to the cluster order parameter
$P_{3D}(p)$ up to the critical zone $II$.%

%TCIMACRO{\FRAME{ftbpFU}{5.8807in}{4.4227in}{0pt}{\Qcb{\QTR{group}{(Color
%online) Plot of spontaneous average magnetization $M$ and specific heat
%$C_{V}$ corresponding to disordered phase ($T>Tc$) and broken phase ($T<Tc$)
%at various probabilities $p$. The continuous curves are a fit to the data
%points. We observe that for large $p\leq1$ values, the behavior $M$ and
%$C_{v}$ is similar to the unbounded case, but with the shift of the critical
%temperature $T_{c}(p)$. }}}{\Qlb{Pic_MCv}}{fig3.eps}%
%width 5.8807in;  height 4.4227in;  depth 0pt;  original-width 5.8219in;
%original-height 4.3708in;  cropleft "0";  croptop "1";  cropright "1";
%cropbottom "0";  filename 'Fig3.eps';file-properties "XNPEU";}}}%
%BeginExpansion
%EndExpansion

Fig.\ref{Pic_MCv} also shows the behavior $M(T)$ and $C_{v}(T)$ in zone $II$
close to the critical point $p_{c3D}\simeq0.311$ (area of incipient
percolation spanning cluster). We observe from Fig.\ref{Pic_MCv} (a), (b) that
separation in the ordered and disordered phases is still seen clearly, and the
critical behavior of $M$ and $C_{v}$ remains well pronounced. However, the
peaks of $C_{v}$ become smoother when compared to zone $I$. Since the
percolation spanning cluster still has a significant number of spins, we can
conclude that in this area the topologically nontrivial magnetic structure
arises; as a result, the critical behavior of $M$ and $C_{v}$ is strongly
affected by the noninteger fractal dimension of the percolation clusters.

Our numerical simulations provide mainly the numerical evidence that the
dependence that the dependence $T_{c}(p)$ is strongly affected by the
noninteger fractal dimension $D_{H}(p)$ of the incipient percolation spanning
cluster in such extended Ising model. More rigorous results relating to the
renormalization group theory and the scaling properties of studied extended
Ising model require significant mathematical efforts and will be published elsewhere.

\section{Conclusion}

In conclusion, we studied the critical properties of Ising spins placed in the
percolation clusters of 3D percolation medium. We focus on a novel aspect of
such a model as an extension of the properties of Ising magnets incorporated
in a percolation medium, where the critical phase transitions would be
affected by the percolation criticality. We found that in such an
inhomogeneous system the standard Ising model is strongly modified by the
spatial percolation cluster distribution. This gives rise to a topologically
nontrivial magnetic structure leading to a modification of the critical
phenomenon without an external magnetic field. We numerically found that at
large occupation probability $p$ (far from criticality) the magnetization
shows well known critical behavior, however, with the shifted critical
temperature $T_{c}(p)$. At a smaller $p$ such dependence is affected by the
noninteger fractal dimension of the incipient percolation spanning cluster. We
discuss analytic formulae that fit the dependencies of the percolation order
parameter and the fractal dimension of clusters on the occupation probability.
Studied features of critical behavior of 3D compound systems (percolation
cluster filled by spins) can open new opportunities in advanced spin- and
optoelectronics of disordered nanostructures, in particular in quantum spin
electronics evolving manipulations of coherent spin states.

\end{document}